\documentclass[%        Class options:
aps,%                   American Physical Society
prd,%                   Physical Review D
showpacs,%              Displays PACS after abstract
preprint,%              Preprint layout
tightenlines,%          Single spaced lines
nofootinbib,%           Does not treat footnotes as references
floatfix]%              Fixes float errors
{revtex4}%              REVTEX 4 Package used
\usepackage{graphicx,%  Default Latex 2eps package for embedding figures
longtable%             Useful for long table
}

\begin{document}
\title{Sterile neutrinos as subdominant warm dark matter}

\author{A.~Palazzo$^{1,2}$, D.~Cumberbatch$^1$, A.~Slosar$^1$, and J.~Silk$^1$}

\address{$^1$ Astrophysics, Denys Wilkinson Building, Keble Road, OX1 3RH,
Oxford, United Kingdom\\
$^2$ Sezione INFN di Bari, Via Amendola 173, 70126, Bari, Italy\\}

%\date{\today}

\begin{abstract}

In light of recent findings which seem to disfavor a scenario with
(warm) dark matter entirely constituted of sterile neutrinos
produced via the Dodelson-Widrow (DW) mechanism, we investigate
the constraints attainable for this mechanism  by relaxing the
usual hypothesis that the relic neutrino abundance must
necessarily account for all of the dark matter. We first study how
to reinterpret the limits attainable from X-ray non-detection and
Lyman-$\alpha$ forest measurements in the case that sterile
neutrinos constitute only a fraction $f_s$ of the total amount of
dark matter. Then, assuming that sterile neutrinos are generated
in the early universe solely through the DW mechanism, we show how
the X-ray and Lyman-$\alpha$ results jointly constrain the
mass-mixing parameters governing their production. Furthermore, we
show how the same data allow us to set a robust upper limit $f_s
\lesssim 0.7$ at the $2 \sigma$ level, rejecting the case of
dominant dark matter ($f_s =1$) at the $\sim 3 \sigma$ level.

\end{abstract}


\maketitle

\section{Introduction}

Recent astrophysical observations, such as the indications of
central cores in low-mass
galaxies~\cite{Dalcanton:2000hn,vandenBosch:2000rz,Swaters:2002rx,Gentile:2004tb,Simon:2004sr,Zackrisson:2006pi,KuziodeNaray:2006wh,Goerdt:2006rw,Strigari:2006ue},
and the low number of satellites observed in Milky Way-sized
galaxies~\cite{Kauffmann:1993gv,Klypin:1999uc,Moore:1999nt},
indicate two possible shortcomings of the [$\Lambda$]CDM paradigm
(see~\cite{Moore:1999gc,Zentner:2003yd}) and have boosted the
interest in a warm dark matter (WDM) scenario which may alleviate
these possible small scale
problems~\cite{Bode:2000gq,Avila-Reese:2000hg}. Electroweak
singlet right handed (``sterile'') neutrinos $\nu_s$ with mass in
the keV range appear to be appealing candidates since they
naturally arise in many
extensions~\cite{Brahmachari:2002va,Berezhiani:1995yi,Chun:1999cq,Langacker:1998ut,Abazajian:2000hw,Asaka:2005an}
of the Standard Model and they could be produced in the early
universe through the Dodelson-Widrow (DW)
mechanism~\cite{Dodelson:1993je} involving (non-resonant)
oscillations with the active species.

{\em Direct} constraints on mass and mixing angle of the sterile
neutrino%
%%%%%%%%%%%%%%%%%%%%%%%%%%%%%%%%%%%%%%%%%%%%%%%%%%%%%%%%%%%%%%
\footnote{Notice that the LSND~\cite{LSND} and
MiniBooNE~\cite{Aguilar-Arevalo:2007it} experiments are not
sensitive to the extremely small mixing angles involved in the DW
mechanism.}
%%%%%%%%%%%%%%%%%%%%%%%%%%%%%%%%%%%%%%%%%%%%%%%%%%%%%%%%%%%%%%
can be obtained by exploiting X-ray
observations~\cite{Abazajian:2001vt,Dolgov:2000ew,Drees:2000wi,Boyarsky:2006fg}.
Indeed, sterile neutrinos posses a radiative decay
channel~\cite{Pal:1981rm,Barger:1995ty}, which gives rise to
photons potentially detectable in astrophysical X-ray and
$\gamma$-ray sources. Limits of such kind have been derived
exploiting diffuse cosmic X-ray
background~\cite{Dolgov:2000ew,Boyarsky:2005us}, Milky Way ``blank
sky''
observations~\cite{Riemer-Sorensen:2006fh,Abazajian:2006jc,Boyarsky:2006ag,Boyarsky:2006hr},
X-ray spectra of nearby
galaxies~\cite{Boyarsky:2006ag,Watson:2006qb} or clusters of
galaxies~\cite{Boyarsky:2006zi,Riemer-Sorensen:2006pi,Boyarsky:2006kc},
and gamma-ray line emission searches in the galactic center
region~\cite{Yuksel:2007xh}. {\em Indirect} constraints on the
neutrino mass can be achieved by the observations of cosmological
structure formation on very small scales, where WDM typically
suppresses the clustering process. In this context,
Lyman-{$\alpha$} forest measurements (henceforth Ly-$\alpha$)
constitute the most suitable probe~\cite{Narayanan:2000tp}, being
sensitive tracers of the primordial density fluctuations on the
smallest scales.

Assuming that sterile neutrinos are produced {\em solely} through
the DW mechanism and requiring that they must account for {\em
all} the dark matter density $\Omega_s = \Omega_{\mathrm {dm}}$,
the recent X-ray
analyses~\cite{Boyarsky:2005us,Riemer-Sorensen:2006fh,Abazajian:2006jc,Boyarsky:2006ag,Boyarsky:2006hr,Watson:2006qb,Boyarsky:2006zi,Riemer-Sorensen:2006pi,Boyarsky:2006kc}
determine upper limits on their mass which lie in the range
(3-8~keV). On the other hand, under the same hypotheses, the
latest Ly-$\alpha$ analyses~\cite{Seljak:2006qw,Viel:2006kd} of
the high redshift flux power spectra, measured by the SDSS
survey~\cite{McDonald:2004eu}, furnish lower bounds in the range
(10-13~keV), in clear tension with the X-ray upper limits.%
%%%%%%%%%%%%%%%%%%%%%%%%%%%%%%%%%%%%%%%%%%%%%%%%%%%%%%%%%%%%%%%%%
\footnote{Similar lower bounds ($\gtrsim 10~\mathrm {keV}$) have
been determined by QSO gravitational lensing
observations~\cite{Miranda:2007rb}.}
%%%%%%%%%%%%%%%%%%%%%%%%%%%%%%%%%%%%%%%%%%%%%%%%%%%%%%%%%%%%%%%%%

While these results imply that sterile neutrinos cannot account
for the {\em entire} amount of dark matter in the universe,%
%%%%%%%%%%%%%%%%%%%%%%%%%%%%%%%%%%%%%%%%%%%%%%%%%%%%%%%%%%%%%%%%%
\footnote{This conclusion can be evaded invoking alternative
production mechanisms (not considered in this work), such as
resonant oscillations~\cite{Shi:1998km}, generation operative
during~\cite{Shaposhnikov:2006xi} or prior
to~\cite{Kusenko:2006rh} oscillations or low reheating
temperatures~\cite{Gelmini:2004ah,Yaguna:2007wi}.}
%%%%%%%%%%%%%%%%%%%%%%%%%%%%%%%%%%%%%%%%%%%%%%%%%%%%%%%%%%%%%%%%%
the possibility exists that the DW mechanism could have produced
only a fraction $f_s = \Omega_s/\Omega_{\mathrm {dm}}$ of its
total content. This does not prevent sterile neutrinos from
playing an important role in other contexts, such as pulsar kicks
~\cite{Kusenko:1997sp,Kusenko:1998bk,Fuller:2003gy,Kusenko:2004mm,Barkovich:2004jp},
supernova explosions~\cite{Hidaka:2006sg,Hidaka:2007}, and
reionization of the
universe~\cite{Barkana:2001gr,Yoshida:2003rm,Mapelli:2005hq,Biermann:2006bu,Mapelli:2006ej,Stasielak:2006br,Ripamonti:2006gr}.
Indeed, it has been already recognized~\cite{Kusenko:2006rh} that
a sterile neutrino explanation of the pulsar kicks can be
reconciled with the existing X-ray constraints assuming the
(lowest possible) relic abundance provided by the DW production
mechanism. Furthermore, the suppression of small scale structures
induced by WDM, while reducing problems at low redshifts, delays
the onset of reionization~\cite{Barkana:2001gr,Yoshida:2003rm},
possibly to an extent difficult to reconcile with the WMAP3
findings~\cite{Spergel:2006hy}. Whether the X-rays emitted by
sterile neutrinos can effectively diminish this inconsistency is
still an open
question~\cite{Biermann:2006bu,Stasielak:2006br,Ripamonti:2006gr},
and a subdominant ($f_s<1$) relic abundance would probably reduce
the tension also in this context.

Therefore, it seems interesting to investigate in a systematic way
the constraints attainable on the mass-mixing parameters which
regulate the DW production mechanism relaxing the usual assumption
of ``dominant'' dark matter ($\Omega_s = \Omega_{\mathrm {dm}}$).
With this purpose in mind, we first address the issue of
reinterpreting the X-ray and Ly-$\alpha$ constraints in the
presence of a $\nu_s$ subdominant relic abundance ($\Omega_s <
\Omega_{\mathrm {dm}}$) which is fixed {\em a priori}, i.e.
independently of the oscillation parameters. Then, we treat the
general case in which the relic abundance depends on mass-mixing
parameters as predicted by the DW mechanism. We determine the
range of such parameters that is compatible with the restrictions
imposed by the joint analysis of X-ray and Ly-$\alpha$
measurements. In addition, we show how these results indirectly
compel the relic abundance to be smaller than $f_s \lesssim 70\%$
(at the $2 \sigma$ level), even when the large (hadronic)
uncertainties affecting the theoretical calculations are taken
into account.

The paper is structured as follows. In Sec.~II, we present our
constraints obtained from the diffuse X-ray background, briefly
discussing their behavior for values of $f_s<1$. In Sec.~III, we
introduce a recipe to derive lower bounds on the sterile neutrino
mass for values $f_s <1$, appropriately rescaling the Ly-$\alpha$
limits already existing for the usual case $f_s = 1$. In Sec.~IV,
we show how the tension existing between X-ray and Ly-$\alpha$
constraints is reduced when the sterile neutrinos are subdominant.
In Sec.~V, we discuss the restrictions imposed on the mass-mixing
parameters by the joint analysis of X-ray and Ly-$\alpha$ data,
also determining the maximum value of the relic abundance allowed
by current data. Finally, we present our conclusions in Sec.~VI.

\section{Constraints from diffuse X-ray data}

Sterile neutrinos possess a (subdominant) radiative decay channel
$\nu_s \to \gamma \nu_a$ into an active neutrino $\nu_a$ and a
photon $\gamma$ with energy equal to a half of the sterile
neutrino mass $E = m_s/2$. For Majorana neutrinos%
% -----------------------------------------------------------------
\footnote{For a Dirac sterile neutrino the decay width is a
half~\cite{Barger:1995ty}.}
% ------------------------------------------------------------------
the radiative decay width can be expressed as%
%%%%%%%%%%%%%%%%%%%%%%%%%%%%%%%%%%%%%%%%%%%%%%%%%%%%%%%%%%%%%%%%%%%
\footnote{Throughout the paper we use natural units with $\hbar =
c = k_B = 1$.}
%%%%%%%%%%%%%%%%%%%%%%%%%%%%%%%%%%%%%%%%%%%%%%%%%%%%%%%%%%%%%%%%%%%
% -----------------------------------------------------------------
\begin{equation}
\Gamma_\gamma = \frac{9 \alpha G_\mathrm{F}^2}{1024 \pi^4} m_s^5
\sin^2 2\theta\ \simeq 1.38 \times 10^{-22} \sin^2 2\theta
\left(\frac{m_s}{\mathrm{keV}}\right)^5 \mathrm{s}^{-1} \,,
\label{eq:Gamma}
\end{equation}
% -----------------------------------------------------------------
where $G_F$ is the Fermi constant, $\alpha$ is the fine-structure
constant and $\theta$ is the mixing angle in vacuum between the
sterile neutrino and the active species which, for the very small
values considered in this work, is related to the effective mixing
angles $\theta_{\alpha}$'s with each of the standard (active)
neutrinos as,
% -----------------------------------------------------------------
\begin{equation}
\sin^2 \theta\ \simeq~\theta^2~ = \displaystyle\sum_{\alpha = e,
\mu, \tau} \theta_{\alpha}^{2} \,. \label{eq:mixing}
\end{equation}
% -----------------------------------------------------------------

In order to derive constraints from the diffuse X-ray background
we use the data provided by the HEAO-1
detector~\cite{marshall:1980,Gruber:1999yr} since this experiment
covers a large energy band, allowing us to put constraints up to
large values of the mass ($\sim 50$~keV). This becomes relevant
when the hypothesis that sterile neutrinos are dominant is
relaxed. Furthermore, these measurements have been shown to
furnish stringent and robust constraints~\cite{Boyarsky:2005us},
which are comparable with those obtained from nearby
galaxies~\cite{Boyarsky:2006ag,Watson:2006qb} or clusters of
galaxies~\cite{Boyarsky:2006zi,Riemer-Sorensen:2006pi,Boyarsky:2006kc}.

The signal expected in the HEAO-1 detector due to the decay of the
putative dark matter candidate originates in part from the extra
galactic (EG) neutrinos, and in part from neutrinos clustered in
the Milky Way (MW) halo. The EG contribution can be evaluated
assuming a uniform distribution of neutrinos in the visible
universe up to very small redshifts. Defining $F_E$ as the present
energy flux of photons produced by neutrino decays, the
differential energy flux (energy flux per unit energy and solid
angle) can be expressed as~\cite{Masso:1999wj}
% -----------------------------------------------------------------
\begin{equation}
\varphi_E^{\mathrm {EG}} \equiv \frac{d^2F^{\mathrm
{EG}}_E}{d\Omega dE} = f_s \frac{\Gamma_\gamma}{4\pi m_s}
\frac{\Omega_{\mathrm {dm}}~\rho_{c}}{H(m_s/2E - 1)}\,,
\label{eq:diff_flux_EG}
\end{equation}
% -----------------------------------------------------------------
where $\rho_c$ is the present critical density, $\Omega_{\mathrm
{dm}}\simeq 0.20$ is its fraction in form of dark
matter~\cite{Spergel:2006hy}, and $H(z)$ is the Hubble function
which, assuming a flat $\Lambda$-matter dominated universe, is
related to the present expansion rate%
%%%%%%%%%%%%%%%%%%%%%%%%%%%%%%%%%%%%%%%%%%%%%%%%%%%%%%%%%%%%%%%%%%%%
\footnote{$H_0$ $\simeq
100~h~\mathrm{km}~\mathrm{s}^{-1}~\mathrm{Mpc}^{-1}$ with $h =
0.73$~\cite{Spergel:2006hy}.}
%%%%%%%%%%%%%%%%%%%%%%%%%%%%%%%%%%%%%%%%%%%%%%%%%%%%%%%%%%%%%%%%%%%%
$H_0$ as
% -----------------------------------------------------------------
\begin{equation}
H(z) \simeq H_0 \sqrt{\Omega_\Lambda + \Omega_{\mathrm {m}} (1 +
z)^3}\,, \label{eq:hubble}
\end{equation}
% -----------------------------------------------------------------
where $\Omega_{\mathrm {m}} \simeq 0.24$ and $\Omega_\Lambda
\simeq 0.76$ are the present fractions of the critical
density~\cite{Spergel:2006hy} in form of (total) matter and dark energy respectively.%
%%%%%%%%%%%%%%%%%%%%%%%%%%%%%%%%%%%%%%%%%%%%%%%%%%%%%%%%%%%%%%%%%%%%
\footnote{We have checked that the uncertainties on the
cosmological parameters have a negligible impact on our results.}
%%%%%%%%%%%%%%%%%%%%%%%%%%%%%%%%%%%%%%%%%%%%%%%%%%%%%%%%%%%%%%%%%%%%

The monochromatic energy flux produced by the sterile neutrinos
clustered in our galactic halo can be formally expressed as
% -----------------------------------------------------------------
\begin{equation}
\varphi_E^{\mathrm {MW}} \equiv \frac{d^2F^{\mathrm
{MW}}_E}{d\Omega dE} = f_s
\frac{\Gamma_\gamma}{8\pi}\frac{m_s}{2E} S_{\mathrm {dm}}
~\delta\left(E-\frac{m_s}{2}\right) \;, \label{eq:diff_flux_MW}
\end{equation}
% -----------------------------------------------------------------
which depends on the direction through the mass column density
$S_\mathrm{dm}$ of dark matter along the line of sight (l.o.s.),
% -----------------------------------------------------------------
\begin{equation}
S_\mathrm{dm} = \int\limits_{\text{l.o.s.}} \rho_{\mathrm
{dm}}(x)dx \;. \label{eq:col}
\end{equation}
% -----------------------------------------------------------------
It is useful to compare the relative magnitude of the EG and MW
contributions through the (directionally-dependent) ratio of their
fluxes integrated over energy. From
Eqs.~(\ref{eq:diff_flux_EG}-\ref{eq:diff_flux_MW}) one gets
% -----------------------------------------------------------------
\begin{equation}
R_{\mathrm {l.o.s}} \equiv \frac{\int \varphi_E^{\mathrm {MW}}dE}{
\int \varphi_E^{\mathrm {EG}}dE} \simeq 0.5 \left(\frac{S_{\mathrm
{dm}}}{10^{-2} {\mathrm g} ~{\mathrm {cm}}^{-2}}\right)\;,
\label{eq:ratio}
\end{equation}
% -----------------------------------------------------------------
which, for typical mass column densities $S_\mathrm{dm}$, turns
out to be of order unity. In particular, adopting the recent
evaluation of the galactic halo mass determined
in~\cite{Battaglia:2005rj} for a Navarro-Frenk-White profile, we
find values of $R$ in the $2\sigma$ range (0.5-3.0), in the
direction of the galactic anti-center (corresponding to the lowest
column density). In consideration of the all-sky coverage of the
HEAO-1 detector~\cite{marshall:1980,Gruber:1999yr}, we adopt the
simple and conservative choice $R = 1$, thus assuming a MW
contribution independent of direction. The effect of a more
realistic treatment (requiring precise knowledge of the sky
coverage of the detector during its time of operation) would only
render our limits more stringent.

In order to find constraints on the mass-mixing parameters, we
have performed a spectral analysis of the HEAO-1 data following
the procedure described in~\cite{Boyarsky:2005us}, adding the EG
and MW contributions to the continuous X-ray spectrum
parameterized by the empirical formula provided
in~\cite{Gruber:1999yr}. In Fig.~1 we show the region excluded at
the $3 \sigma$ level (above the curves) obtained in the case $f_s
= 1$. For completeness, we show separately the constraints that we
obtain including {\em only} the EG contribution (dotted line), and
{\em only} the MW contribution (dashed line), together with the
constraints obtained properly taking both contributions into
account (solid line). Even if the total MW and EG fluxes have the
same order of magnitude (exactly identical in our case), the MW
contribution has a sensibly larger impact in determining the
constraints. This different sensitivity is due to the different
form of the energy spectrum produced in the two cases. Indeed,
while the MW signal is just a line broadened into a (large)
gaussian (centered around $E = m_s/2$) by the (poor) energy
resolution ($\Delta E/E \simeq 25\%$) of the
detector~\cite{Gruber:1999yr}, the EG spectrum is intrinsically
broadened towards low energies as a result of the integration over
redshift and is further enlarged by the effect of the resolution
of the detector. Both constraints obtained with or without the
inclusion of the MW contribution are in good agreement with the
analogous ones found respectively in~\cite{Boyarsky:2005us}
and~\cite{Boyarsky:2006fg}. Our bounds appear only slightly weaker
when  the MW contribution is also included, presumably due to the
different choice adopted for the mass of the galactic halo.

As is evident from Eqs.~(\ref{eq:Gamma}-\ref{eq:diff_flux_MW}),
the differential flux produced by the decay of sterile neutrinos
depends on the product $f_s \sin^2 2\theta$. Hence, for a fixed
neutrino mass, the upper limit on the mixing angle must weaken for
decreasing values of $f_s$, rescaling exactly as the inverse of
the fraction $f_s$. This means that if one fixes {\em a priori}
the fraction $f_s<1$ (i.e. independently of the mass-mixing
parameters), the excluded region in the plane $[m_s,
\sin^2{2\theta}]$ will reproduce that found for the usual case
$f_s =1$ modulo a ``rigid'' shift towards larger values of the
mixing angle (see Sec.~IV). The form of the contours will be
altered in a different way when, as predicted by the DW mechanism,
the relic abundance depends on the mass-mixing parameters (see
Sec.~V).

\section{Constraints from the Lyman-$\alpha$ forest}

Massive neutrinos affect the cosmological evolution through two
distinct effects. Firstly, in the early universe, they are
relativistic and therefore contribute to the overall radiation
budget of the universe, shifting the epoch of matter-radiation
equality. Secondly, neutrinos erase fluctuations on scales smaller
than the horizon size at the epoch when their kinematics becomes
non-relativistic -- the so-called free-streaming
length~\cite{Bond:1980ha}. For neutrinos under discussion here,
the first effect is completely negligible while the second one can
leave important imprints on structure formation which are
detectable by Ly-$\alpha$ forest measurements. Indeed, these
observations are directly sensitive to the free-streaming length
which can be roughly expressed as~\cite{Abazajian:2001nj}
% -----------------------------------------------------------------
\begin{equation}
\lambda_{\mathrm {FS}} \simeq 1.2~{\mathrm {Mpc}}
\left(\frac{\mathrm {keV}}{m_s} \right) \frac{\langle p/T
\rangle}{3.15} \;, \label{eq:freestream}
\end{equation}
% -----------------------------------------------------------------
where $\langle p/T \rangle$ is the mean momentum over temperature
for the sterile neutrino distribution. To impose constraints at
such low length scales ($\lesssim \mathrm {Mpc}$), one needs a
probe of the small-scale matter power spectrum at high redshift,
where the information on the primordial fluctuations has not yet
been completely lost due to non-linear evolution. The Ly-$\alpha$
forest at redshifts 2-4 satisfies both requirements, constituting
the most suitable probe of WDM~\cite{Narayanan:2000tp}. Two recent
and independent Ly-$\alpha$ analyses have provided very stringent
lower bounds on the neutrino mass. Assuming that {\em all} the
dark matter is made of sterile neutrinos and that they are
produced via non-resonant oscillations, a lower bound of $13$~keV
(at $95\%$~C.L.) has been determined in~\cite{Seljak:2006qw}, and
a slightly weaker limit of $10$~keV (at $95\%$~C.L.) has been
obtained in~\cite{Viel:2006kd}.

The most appropriate approach to constrain mixed models with
$f_s<1$ (we assume that the remaining fraction $(1 - f_s)$ of dark
matter is made of CDM) would be by running a grid of
hydrodynamical simulations in order to relate the observed
Ly-$\alpha$ flux power spectrum with the expanded set of
parameters considered here. This exceeds the scope of this work,
and we instead adopt a different method which allows us to obtain
reliable constraints simply rescaling the ones found
in~\cite{Seljak:2006qw} for the case $f_s=1$.

Assuming fiducial values for all cosmological parameters, we
calculate the growth of perturbations including a mixed
contribution of WDM ($f_s$) and of CDM ($1-f_s$) in the code
CAMB~\cite{CAMB}, incorporating the $\nu_s$ relic abundance
through the relation
% -----------------------------------------------------------------
\begin{equation}
\Omega_s h^2 = \beta \left(\frac{m_s}{93.2 \mathrm {eV}}\right)
\;, \label{eq:density}
\end{equation}
% -----------------------------------------------------------------
where $\beta$ is a suppression factor and we assume that sterile
neutrinos posses a thermal momentum distribution. Recent
calculations~\cite{Abazajian:2005gj,Asaka:2006rw} have shown that
the real distribution exhibits appreciable deviations from the
thermal form, which can be roughly approximated by a moderate
shift of the average momentum toward lower
values~\cite{Abazajian:2005gj,Asaka:2006rw}. This implies [see
Eq.~(\ref{eq:freestream})] smaller free-streaming lengths and a
consequent weakening of the lower bounds on the mass derived with
the assumed approximation of a thermal distribution. The limit of
$13$~keV quoted in~\cite{Seljak:2006qw} already accounts for a
$10\%$ correction ($\langle p/T  \rangle \sim 0.9$) as evaluated
in~\cite{Abazajian:2005gj}. Since the
calculations~\cite{Asaka:2006rw} indicate deviations as large as
$\simeq 20\%$ ($\langle p/T \rangle \sim 0.8$ for masses of
10~keV), we have conservatively rescaled the lower bound obtained
in~\cite{Seljak:2006qw} by a further factor $0.9$, thus adopting
for the pure WDM  case ($f_s = 1$) the reference lower bound $m_s
\gtrsim 11.5~\mathrm {keV}$.

Due to their pencil-beam nature, Ly-$\alpha$ measurements
effectively probe the projected 1D flux power spectrum rather than
the full 3D power spectrum. This implies that small-scale modes
are projected onto much larger modes~\cite{Seljak:2006qw}, with
enhanced sensitivity on the scales of interest. Therefore, for
each model (defined by the values of $f_s$ and $m_s$), we
integrate the resulting 3D linear power spectrum into the relevant
1D spectrum, using the relation
% -----------------------------------------------------------------
\begin{equation}
P_{\mathrm {1D}}(k) = \frac{1}{2\pi} \int_k^\infty P_{\mathrm
{3D}}(k) k dk \;. \label{eq:P1D}
\end{equation}
% -----------------------------------------------------------------

In order to derive lower limits on the neutrino mass from those
already calculated in~\cite{Seljak:2006qw} for the case of pure
WDM ($f_s = 1$), we adopt the following rescaling
procedure~\cite{SeljakPrivate}. For a model with a WDM fraction
$f_s<1$, we find the value of the neutrino mass that produces a
suppression of the 1D power spectrum at the fiducial pivot
wavenumber $k_f = 2h$~Mpc$^{-1}$ equal to that one obtained (for a
larger value of the mass) in the case $f_s =1$. Once the
confidence level (C.L) at which a given mass is disfavored in the
pure WDM case is provided, we assume that the values of the mass
calculated as above for the models with $f_s<1$ are disfavored at
the same C.L.

Our procedure can be visualized in Fig.~\ref{fig:supp_2mpc} which
displays in the plane [$m_s$, $f_s$] the iso-contours of the
fractional suppression of the 1D matter power spectrum obtained
for mixed (CDM +WDM) models with respect to the case of pure CDM.
The values refer to the pivot scale $k = 2h$~Mpc$^{-1}$. The
region under the thick solid line is excluded at the $2 \sigma$
level and corresponds to a fractional suppression of $\sim 23\%$,
which is the value found in~\cite{Seljak:2006qw} for the case of
pure WDM  corresponding to the quoted $2\sigma$ lower bound on the
neutrino mass. We observe that for small values of the relic
abundance the lower limits depart from the power-law behavior
displayed at large values, indicating a loss of sensitivity of the
Ly-$\alpha$ measurements.

Although we have chosen the pivot wavenumber $k_f=2h$~Mpc$^{-1}$
as this is the scale where SDSS Ly-$\alpha$ data are most
sensitive~\cite{Seljak:2006qw}, we have checked that our rescaling
procedure returns stable results with respect to possible
different choices of the pivot wavenumber in the range
(1-5$~h$~Mpc$^{-1}$). In fact, a different choice of $k_f$ in such
range alters the lower limits by less than $10\%$, provided that
sufficiently high values of the WDM fraction ($f_s \gtrsim 0.1$)
are considered. For smaller values of $f_s$, the discrepancies
become more pronounced and our procedure should be replaced by a
more accurate treatment involving detailed simulations. We also
observe that our rescaling procedure tacitly assumes that the
change of cold to warm dark matter fraction affects just the
linear power spectrum while leaving the transfer functions from
the linear 1D power spectrum to the observed flux power spectrum
unchanged. Corrections to this simple approximation could arise
since the 1D power spectrum probes very high $k$ modes which are
probably significantly affected by the Jeans smoothing, non-linear
coupling from larger scales and redshift-space
distortions~\cite{McDonaldPrivate}. Therefore, dedicated hydrodynamical simulations are required in
order to render our conclusions more quantitative.

\section{Theoretical predictions and experimental constraints}

In this section we will show how the constraints coming from X-ray
and Ly-$\alpha$ data relate to the theoretical restrictions
imposed by non-resonant production by confronting the usual case
of dominant dark matter ($f_s = 1$) with a representative case in
which it is subdominant ($f_s = 0.2$). The results of this
sections will serve also as a guide for the interpretation of the
more general results presented in the next section.

For our purposes we use the theoretical calculations performed
in~\cite{Asaka:2006rw}. Although similar calculations have been
recently performed~\cite{Abazajian:2005gj}, which are in
substantial agreement with those presented in~\cite{Asaka:2006rw},
the latter incorporate the most conservative evaluation of the
hadronic uncertainties affecting the production process.
Furthermore, fitting formulae valid in the whole range of mass and
mixing parameters explored here are provided which allow us draw
iso-abundance curves in the mass-mixing plane and to incorporate
the theoretical predictions in the analysis. According to the
calculations performed in~\cite{Asaka:2006rw}, after a slight
change in the notation, the relic abundance of sterile neutrinos
can be conveniently expressed as
% -----------------------------------------------------------------
\begin{equation}
\Omega_s h^2 \simeq 0.275~ \left(\frac{\sum_{\alpha}
C_\alpha(m_s)\,\theta^2_{\alpha }}{\sum_{\alpha}
\theta^2_{\alpha}} \right)\left(\frac{m_s}{\mathrm
{keV}}\right)^{2} \left( \frac{\sin^2 2\theta}{10^{-7}}\right) \;,
\label{eq: Omega_s}
\end{equation}
% -----------------------------------------------------------------
where the first term in parentheses embeds a mild dependence on
the neutrino mass and on the flavor structure%
%%%%%%%%%%%%%%%%%%%%%%%%%%%%%%%%%%%%%%%%%%%%%%%%%%%%%%%%%%%%%%%%
\footnote{The functions $C_\alpha$ exhibit a moderate
hierarchy~\cite{Asaka:2006rw} ($C_e > C_\mu > C_\tau$).}
%%%%%%%%%%%%%%%%%%%%%%%%%%%%%%%%%%%%%%%%%%%%%%%%%%%%%%%%%%%%%%%%
through the slowly varying functions $C_\alpha$, which are
directly evaluated in~\cite{Asaka:2006rw}. From direct comparison
of Eqs.~(\ref{eq:density}) and (\ref{eq: Omega_s}), we observe
that the suppression factor $\beta$ in Eq.~(\ref{eq:density}) is
related to the production mechanism. In particular, once the relic
abundance and the mass are fixed, the factor $\beta$ is
unequivocally determined by the mixing angles. This dependence,
together with the deviations from the thermal distribution,
distinguishes the case of (out-of-equilibrium) production via
non-resonant oscillations from the case of early decoupling (in
equilibrium), where an analogous suppression factor is instead
related to the number of relativistic degrees of freedom at the
time of decoupling~\cite{Colombi:1995ze}.

The fitting formulae presented in~\cite{Asaka:2006rw} provide the
relation that mass and mixing parameters must satisfy if sterile
neutrinos constitute all the amount of dark matter. These are
derived imposing $\Omega_s = \Omega_{\mathrm {dm}}$ in
Eq.~(\ref{eq: Omega_s}), and thus correspond to the case $f_s =
1$. Given the form of Eq.~(\ref{eq: Omega_s}), one can obtain
fitting formulae for the fractional relic abundance simply
rescaling those ones valid for the case $f_s = 1$. In particular,
for the best estimate of the hadronic effects, one gets the average%
%%%%%%%%%%%%%%%%%%%%%%%%%%%%%%%%%%%%%%%%%%%%%%%%%%%%%%%%%%%%%
\footnote{Here, we have taken the average of the two cases
respectively dubbed as ``case 1, mean"  and ``case 2, mean"
in~\cite{Asaka:2006rw}, obtained for two extreme flavor
structures.}
%%%%%%%%%%%%%%%%%%%%%%%%%%%%%%%%%%%%%%%%%%%%%%%%%%%%%%%%%%%%%
abundance~\cite{Asaka:2006rw}
%-----------------------------------------------------------------
\begin{equation}
\log_{10} \left(f_s^{\mathrm {ave}}\right) = +0.17 +
1.84~\log_{10}\left(\frac{m_s}{\mathrm {keV}} \right) +
\log_{10}\left(\frac{\sin^2 2\theta}{10^{-7}} \right) \;,
\label{eq:fs_ave}
\end{equation}
%-----------------------------------------------------------------
which we will use as the reference estimate in our analysis.
Taking into account the hadronic uncertainties, the authors
of~\cite{Asaka:2006rw} determine two extreme cases, corresponding
respectively to the minimal
%-----------------------------------------------------------------
\begin{equation}
\log_{10} \left(f_s^{\mathrm {min}}\right) = -0.07 +
1.74~\log_{10}\left(\frac{m_s}{\mathrm {keV}} \right) +
\log_{10}\left(\frac{\sin^2 2\theta}{10^{-7}} \right) \;,
\label{eq:fs_min}
\end{equation}
%-----------------------------------------------------------------
and to the maximal
%-----------------------------------------------------------------
\begin{equation}
\log_{10} \left(f_s^{\mathrm {max}}\right) = +0.62 +
1.74~\log_{10}\left(\frac{m_s}{\mathrm {keV}} \right) +
\log_{10}\left(\frac{\sin^2 2\theta}{10^{-7}} \right) \;,
\label{eq:fs_max}
\end{equation}
%-----------------------------------------------------------------
abundances that can be produced for given mass-mixing parameters.
These estimates constitute two extreme cases in that they have
been obtained by ``pushing" all the errors induced by different
and in principle independent sources of uncertainty%
%%%%%%%%%%%%%%%%%%%%%%%%%%%%%%%%%%%%%%%%%%%%%%%%%%%%%%%%%%%%%%%%%%
\footnote{These are mainly the uncertainties on the
equation-of-state (EOS) and on the hadronic scattering
processes~\cite{Asaka:2006rw}.}
%%%%%%%%%%%%%%%%%%%%%%%%%%%%%%%%%%%%%%%%%%%%%%%%%%%%%%%%%%%%%%%%%%%
towards the same direction. Lacking a precise definition of their
statistical significance, we will adopt the conservative choice to
interpret them as $2 \sigma$ limits in our analysis.

In order to show how the constraints imposed by X-ray and
Ly-$\alpha$ data relate to the theoretical predictions, we
superimpose in Fig.~\ref{sections} all the ($2 \sigma$)
experimental restrictions obtained for the two fixed values $f_s =
1$ (left panel) and $f_s = 0.2$ (right panel) to the theoretical
curves corresponding to these same values and obtained by
Eqs.~(\ref{eq:fs_ave}-\ref{eq:fs_max}). In both panels, the region
above the descending thick solid curve is excluded by the X-ray
analysis while the region below the thick horizontal line is
forbidden by Ly-$\alpha$. The inclined dashed line corresponds to
the best estimate given by Eq.~(\ref{eq:fs_ave}), while the other
two dotted lines represent the two extreme cases of
Eqs.~(\ref{eq:fs_min}-\ref{eq:fs_max}).

In the left panel, corresponding to the usual case $f_s = 1$ (see
Refs.~\cite{Watson:2006qb,Viel:2006kd} for analogous plots), the
strong tension existing between experimental constraints and
theoretical predictions is clearly evidenced: the theoretical
curves lie in the region which is excluded by X-ray data, by
Ly-$\alpha$ data or by both sets of data. In particular, looking
at the intersections points between the theoretical curve and the
border of the regions excluded by X-ray and Ly-$\alpha$ data one
find the incompatible limits $m_s \lesssim 7$~keV and $m_s \gtrsim
11.5$~keV respectively. Hence, as expected, the dominant case $f_s
= 1$ seems strongly disfavored.

In the right panel, all the curves (including the theoretical one)
refer to fractional abundance $f_s = 0.2$. Here the iso-abundance
lines move (for a fixed mass) toward noticeably smaller values of
the mixing angle. On the contrary, as discussed in Sec.~II, the
X-ray limits become less stringent ``moving" toward larger (by a
factor $1/f_s = 5$) values of the mixing angle, while the
Ly-$\alpha$ lower bound on the mass drastically decreases ($m_s
\gtrsim 2.5$~keV) as discussed in Sec.~III. The net result is that
for subdominant dark matter with fractional abundance $f_s=0.2$ a
region for the mass-mixing parameters is permitted. If one adopts
the best estimate for the relic abundance, this region corresponds
to the segment of the dashed line delimited by the two regions
excluded respectively by X-ray and Ly-$\alpha$ data. The end
points of such segment correspond to values of the neutrino mass
in the range (2.5-16~keV) and to values of the mixing angle in the
range ($9 \times 10^{-11}$--$2.5\times 10^{-9}$). Intuitively one
can expect that such ranges would change (reducing) for increasing
values of the fractional abundance $f_s$ until they disappear for
sufficiently high values of $f_s$. This is the case as we will
show by the quantitative analysis performed in the next section.

\section{Constraints on the Dodelson-Widrow scenario}

In this section we will show how the results coming from X-ray and
Ly-$\alpha$ data can be used to put constraints on the parameters
which govern non-resonant production of sterile neutrinos in the
early universe. In order to determine the mass-mixing parameters
which are compatible with the experimental data, we impose that
for each point of the mass-mixing plane, the relic abundance is
the one determined by the theoretical calculations.

Figure~\ref{fig:qcd_impact} shows the results of such procedure
for the three different estimates provided by
Eqs.~(\ref{eq:fs_ave}-\ref{eq:fs_max}). The first panel
corresponds to the ``average" case [Eq.~(\ref{eq:fs_ave})], while
the second and third panels represent respectively the two extreme
cases of lowest [Eq.~(\ref{eq:fs_min})] and highest
[Eq.~(\ref{eq:fs_max})] production efficiencies. In each of the
three panels the solid curve delimits the parameter region (below)
allowed at $2 \sigma$ level, while the dotted lines represent
iso-abundance contours. The left upper branch of the curve is
determined by the X-ray constraints which now display a different
behavior with respect to those presented in Fig.~\ref{sections}.
Indeed, the relic abundance is not fixed {\em a priori} but varies
with the parameters. Consequently, the limits on the neutrino mass
now decrease faster with the mixing angle due to the corresponding
increase of the relic abundance along the curve. As already
observed in~\cite{Kusenko:2006rh} these limits are less stringent
than those obtained for the case $f_s = 1$ (see the left panel of
Fig.~\ref{sections}), thus leaving room for a sterile neutrino
explanation of the pulsar kicks. The allowed region is limited on
the right by the fast descending branch determined by Ly-$\alpha$
measurements which, fixing a lower bound on the mass, indirectly
limits also the mixing angle. The two branches converge for
increasing values of $f_s$ ``closing'' the allowed region in
correspondence of a maximum value of the relic abundance
(evidenced by a dashed segment) which depends on the assumed
theoretical calculation. We observe that the left upper branch of
the allowed region in the top panel of Fig.~\ref{fig:qcd_impact}
is similar to that one determined in~\cite{Kusenko:2006rh} using
the same procedure. However, since only the X-ray data are
considered in~\cite{Kusenko:2006rh}, the right lower branch of the
allowed region is determined in~\cite{Kusenko:2006rh} by
the``overclosure'' condition ($f_s =1$), which is less restrictive
than the constraints we obtain including also the Ly-$\alpha$
data. Moreover, we stress that only the inclusion of both set of
data (X-ray and  Ly-$\alpha$) is able to jointly constrain the
fractional abundance. The comparison of the three panels of
Fig.~\ref{fig:qcd_impact} allows a qualitative evaluation of the
impact of the theoretical uncertainties. In particular, the
allowed region is subject to appreciable changes, enlarging
(reducing) for lower (higher) production efficiencies and the
upper limit on the abundance is subject to noticeable fluctuations
in the range (0.55-0.75).

In order to incorporate such uncertainties in a quantitative way
in our analysis we allow the fractional abundance $f_s$ to vary
around its average value [determined by Eq.~(\ref{eq:fs_ave})]
according to a log-normal%
%%%%%%%%%%%%%%%%%%%%%%%%%%%%%%%%%%%%%%%%%%%%%%%%%%%%%%%%%%%%%%%%%%%
\footnote{The log-normal distribution is more suited to treat the
large hadronic uncertainties.}
%%%%%%%%%%%%%%%%%%%%%%%%%%%%%%%%%%%%%%%%%%%%%%%%%%%%%%%%%%%%%%%%%%%
distribution with standard deviation taken equal to one half of
the excursion (respect to the average) determined by the two
extreme cases in Eqs.~(\ref{eq:fs_min}-\ref{eq:fs_max}). This
corresponds to adding a penalty factor%
%%%%%%%%%%%%%%%%%%%%%%%%%%%%%%%%%%%%%%%%%%%%%%%%%%%%%%%%%%%%%%%%%%
\footnote{Note that the three cases considered above (in which the
relic abundance is univocally fixed by the mass and mixing
parameters) are formally recovered introducing as penalty factor a
Dirac delta function centered respectively on the average, lowest
or highest abundance.}
%%%%%%%%%%%%%%%%%%%%%%%%%%%%%%%%%%%%%%%%%%%%%%%%%%%%%%%%%%%%%%%%%%
$\eta$ to the total $\chi^2$
%-----------------------------------------------------------------
\begin{equation}
\chi^2 = \chi^2_{\mathrm {X-ray}}(\sin^2 2\theta, m_s, f_s) +
\chi^2_{\mathrm{Ly-\alpha}}(m_s, f_s) + \eta(\sin^2 2\theta, m_s,
f_s)  \;, \label{eq:chi2}
\end{equation}
%-----------------------------------------------------------------
defined as
%-----------------------------------------------------------------
\begin{equation}
\eta = \left[\frac{\log_{10} (f_s) - \log_{10}(f_s^{\mathrm
{ave}})} {\Delta \log_{10} (f_s)}\right]^{2} \;,
\label{eq:penalty}
\end{equation}
%-----------------------------------------------------------------
with $1 \sigma$ (asymmetric) errors given by
%-----------------------------------------------------------------
\begin{equation}
\Delta \log_{10}(f_s) = 0.5~[\log_{10}(f_s^{\mathrm {max}}) -
\log_{10}(f_s^{\mathrm {ave}})] ~~~~~~~(f_s > f_s^{\mathrm {ave}})
\;, \label{eq:err1}
\end{equation}
%-----------------------------------------------------------------
%-----------------------------------------------------------------
\begin{equation}
\Delta \log_{10}(f_s) = 0.5~[\log_{10}(f_s^{\mathrm {ave}}) -
\log_{10}(f_s^{\mathrm {min}})]  ~~~~~~~(f_s < f_s^{\mathrm
{ave}}) \;. \label{eq:err2}
\end{equation}
%-----------------------------------------------------------------
Then, in order to obtain the allowed region for the mass-mixing
parameters, we marginalize the $\chi^2$ in Eq.~(\ref{eq:chi2})
with respect to the parameter $f_s$. Figure~\ref{fig:mass_mixing}
shows the allowed region (at the 2$\sigma$ and 3$\sigma$ levels)
obtained following this procedure. As expected, the $2 \sigma$
region is slightly enlarged respect to that one obtained for the
``average'' case (first panel of Fig.~\ref{fig:qcd_impact}). Note
that in this plot iso-abundance contours cannot be drawn since the
value of $f_s$ is now determined by the marginalization process
and thus depends on the confidence level. In particular, the
maximum value allowed at the $2 \sigma$ level turns out to be
$f_s\lesssim 0.7$ which is slightly more stringent than the
absolute upper limit $(0.75)$ obtained for the extreme case of the
highest production efficiency (third panel of
Fig.~\ref{fig:qcd_impact}). In fact, any departure from the best
estimated abundance is now hindered by the counterbalancing effect
of the penalty factor $\eta$. Furthermore, we find that the case
in which sterile neutrinos constitute all the amount of dark
matter ($f_s = 1$) is disfavored at the $\sim 3\sigma$ level.

As already noted in the previous section (see the discussion
concerning the case $f_s = 0.2$), for a given value of the
relative abundance $f_s$ produced via non-resonant oscillations,
the experimental data determine two allowed ranges for the mass
and the mixing angle respectively. From Fig.~\ref{fig:qcd_impact}
one can see that the amplitude of such ranges gradually diminishes
until it eventually becomes zero for the maximum value allowed for
the abundance. This behavior can be effectively visualized
marginalizing the $\chi^2$ in Eq.~(\ref{eq:chi2}) with respect to
one of the mass-mixing parameters. The result of such an exercise
is shown in Fig.~\ref{fig:ranges} which shows the (2$\sigma$ and
3$\sigma$) ranges for values of $f_s>0.1$, where our Ly-$\alpha$
rescaling procedure furnishes robust results.%
%%%%%%%%%%%%%%%%%%%%%%%%%%%%%%%%%%%%%%%%%%%%%%%%%%%%%%%%%%%%%%%%%%%
\footnote{It has been recently shown~\cite{Yuksel:2007xh} that
gamma-ray line emission searches can improve current X-ray limits
in the region of very large masses and small mixing angles, which
is relevant for very low values of the relic abundance
($f_s<0.1$).}
%%%%%%%%%%%%%%%%%%%%%%%%%%%%%%%%%%%%%%%%%%%%%%%%%%%%%%%%%%%%%%%%%%%
In the first panel (where the mass is projected out), the lower
bound on the mixing angle is determined by X-ray observations,
while the upper bound is fixed by Ly-$\alpha$ measurements. The
situation is reversed in the second panel (here the mixing angle
is projected out) where the range allowed for the neutrino mass is
limited from below by Ly-$\alpha$ measurements, and the upper
bound is provided by X-ray observations.

Although a direct measurement of the relic abundance of sterile
neutrinos is not feasible, an indirect evidence for a mixed
(WDM+CDM) scenario may constitute a realistic possibility. In this
case, one might ask which parameters can give rise to sterile
neutrinos in the right amount required to make the WDM component.
Figure~\ref{fig:ranges} provides an immediate and quantitative
answer to this question showing at a glance where the DW model
should ``live''. As expected the allowed ranges gradually reduce
for increasing values of $f_s$ until they eventually shrink in
correspondence of the maximum value allowed for the relic
abundance.

We close this section with a final cautionary remark. As already
stressed, the upper limit that we found on the relic abundance is
valid only under the assumption that sterile neutrinos are
produced {\em solely} through the simplest production mechanism
involving non-resonant oscillations. It can be evaded considering
exotic models such as resonant oscillations~\cite{Shi:1998km}
boosted by large primeval lepton asymmetries, generation
mechanisms operative during~\cite{Shaposhnikov:2006xi} or prior
to~\cite{Kusenko:2006rh} oscillations or low reheating
temperatures~\cite{Gelmini:2004ah,Yaguna:2007wi}. Therefore,
further investigation of these possibilities is important to
prevent a viable dark matter candidate from being dismissed
prematurely. In this context our results could serve as a
quantitative guide for exploring possible mechanisms which are
expected to act in synergy with the production via non-resonant
oscillations.

%.....................................
\section{Conclusions}
%.....................................

In light of recent results which seem to disfavor keV sterile
neutrinos as viable dark matter candidates, we have revisited the
constraints attainable on the parameters which govern the
Dodelson-Widrow (DW) production mechanism. Relaxing the usual
hypothesis that sterile neutrinos must account for all the dark
matter content ($\Omega_s = \Omega_{\mathrm {dm}}$), we have shown
how the X-ray and Ly-$\alpha$ measurements can be reinterpreted in
the subdominant case ($\Omega_s < \Omega_{\mathrm {dm}}$). In
addition, we have shown how the current data provide us with a
conservative upper bound on the fraction of sterile neutrinos
produced via the DW mechanism, which is robust with respect to the
large uncertainties affecting the theoretical estimates. More
sensitive X-ray observations, more quantitative Ly-$\alpha$
analyses, and a reduction of the theoretical uncertainties, all
will need to play a crucial role in order to improve our limits.

\acknowledgments

We are grateful to V. Antonuccio, E.~Lisi, P.~McDonald,
D.~Montanino, S.~Sarkar, and U.~Seljak for useful discussions.
A.~Palazzo is supported by INFN, A.~Slosar by Oxford Astrophysics.

%%%%%%%%%%%%%%%%%%%%%%%%%%%%%%%%%%%%%%%%%%%%%%%%%%%%%%%%%%%%%%%%%%%%%%%%%
\begin{figure}
\includegraphics[scale=0.90, bb= 100 100 510 720]{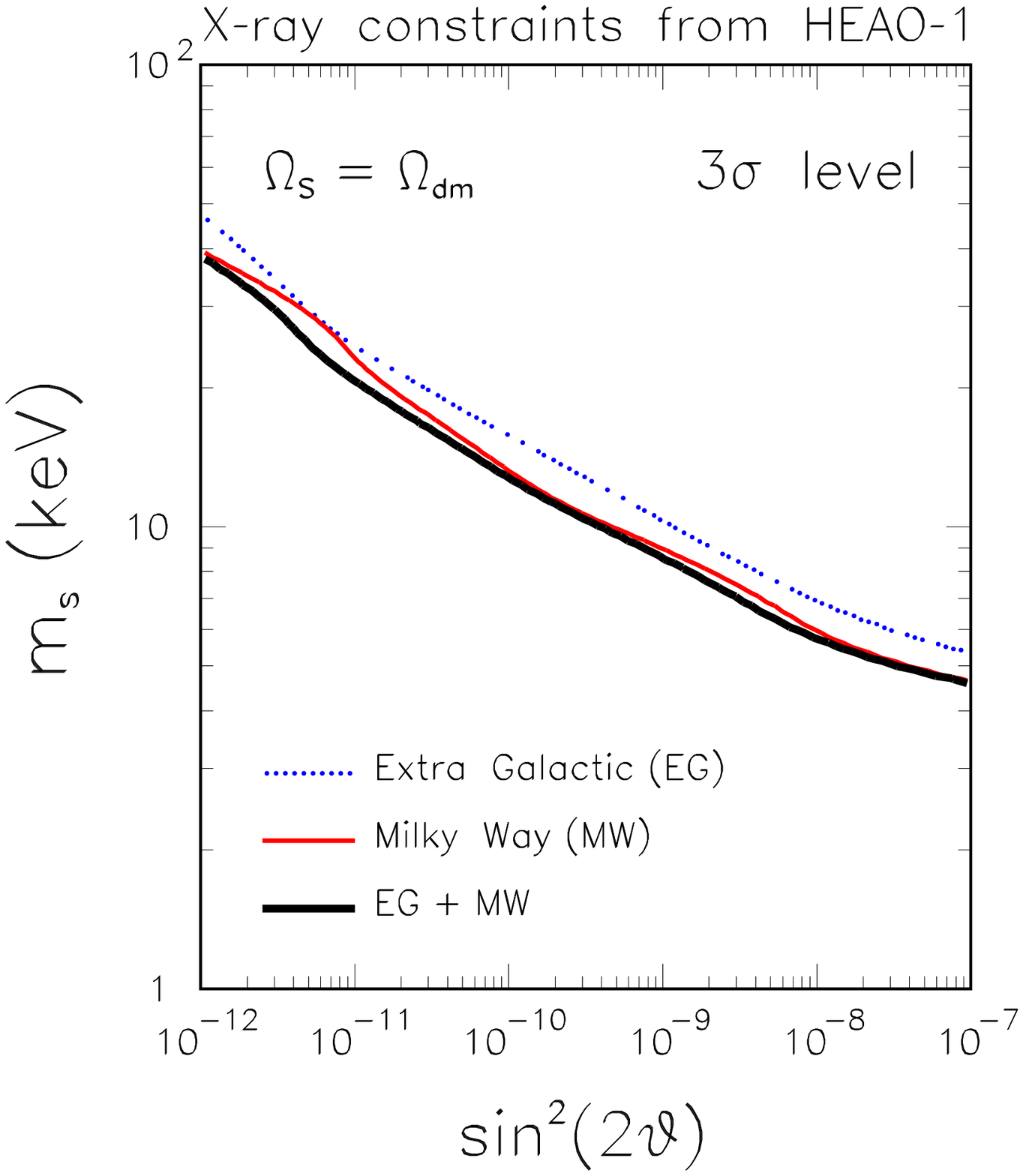}
\vspace*{0.0cm} \caption{ \label{fig:xray} Constraints (at the $3
\sigma$ level) on mass and mixing parameters obtained from the
spectral analysis of the HEAO-1 data under the assumption that
sterile neutrinos account for all the dark matter content
($\Omega_s = \Omega_{\mathrm {dm}}$). The bounds obtained
including only the Milky Way (MW) contribution, only the
extra-galactic (EG) contribution, and their sum are shown
separately.}
\end{figure}
%%%%%%%%%%%%%%%%%%%%%%%%%%%%%%%%%%%%%%%%%%%%%%%%%%%%%%%%%%%%%%%%%%%%%%%%%

%%%%%%%%%%%%%%%%%%%%%%%%%%%%%%%%%%%%%%%%%%%%%%%%%%%%%%%%%%%%%%%%%%%%%
\begin{figure}
\includegraphics[scale=0.90, bb= 100 100 510 720]{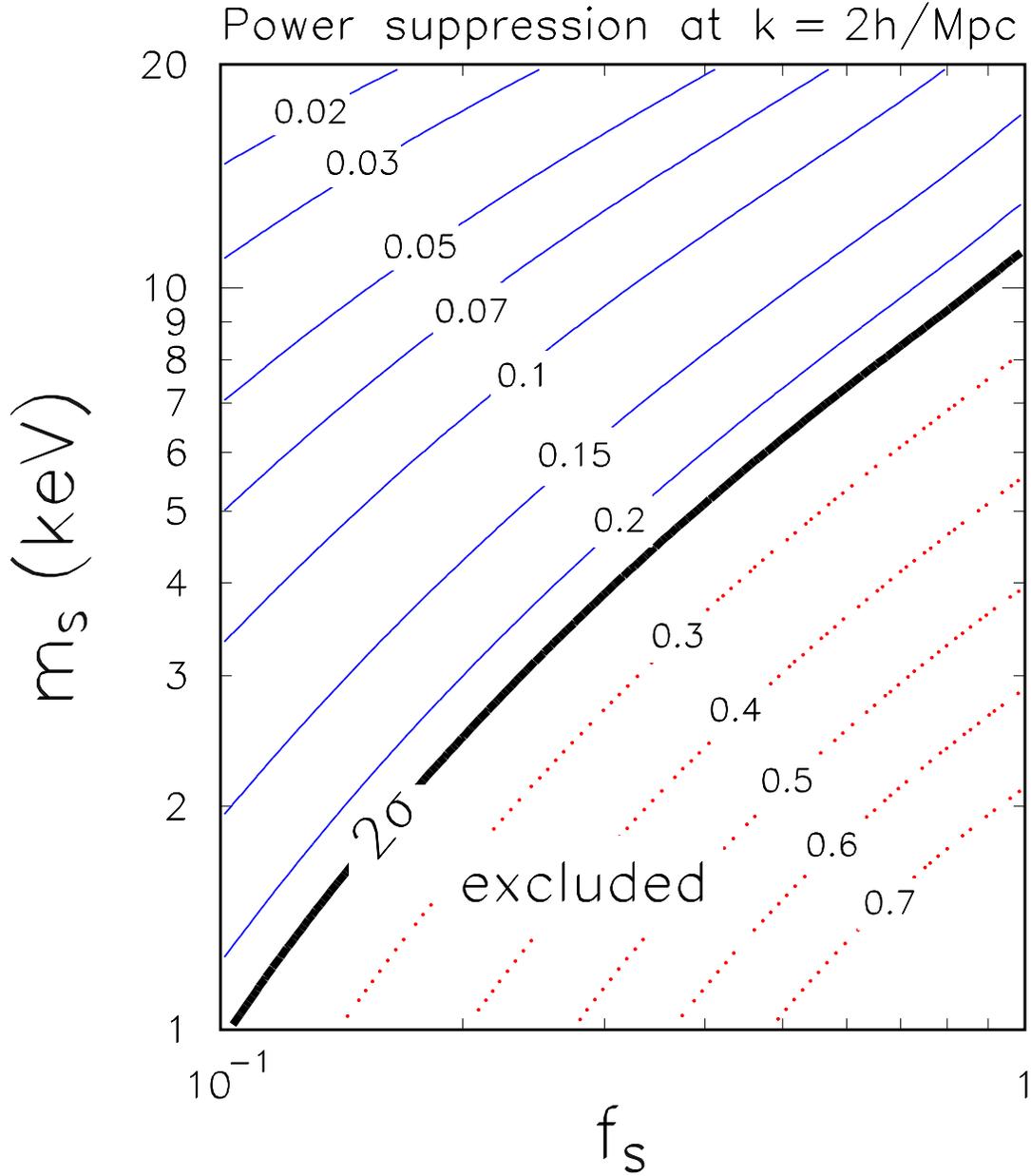}
\vspace{0.0cm}\caption{\label{fig:supp_2mpc} Fractional
suppression of the one dimensional matter power spectrum for mixed
(CDM +WDM) models with respect to the case of pure CDM. The values
refer to the pivot scale $k = 2h$~Mpc$^{-1}$. The region under the
thick solid line is excluded at $2 \sigma$ level by Lyman-$\alpha$
measurements. See the text for details.}
\end{figure}
%%%%%%%%%%%%%%%%%%%%%%%%%%%%%%%%%%%%%%%%%%%%%%%%%%%%%%%%%%%%%%%%%%%%%

%%%%%%%%%%%%%%%%%%%%%%%%%%%%%%%%%%%%%%%%%%%%%%%%%%%%%%%%%%%%%%%%%%%%%
\begin{figure}
\includegraphics[scale=0.90, bb= 100 100 510 720]{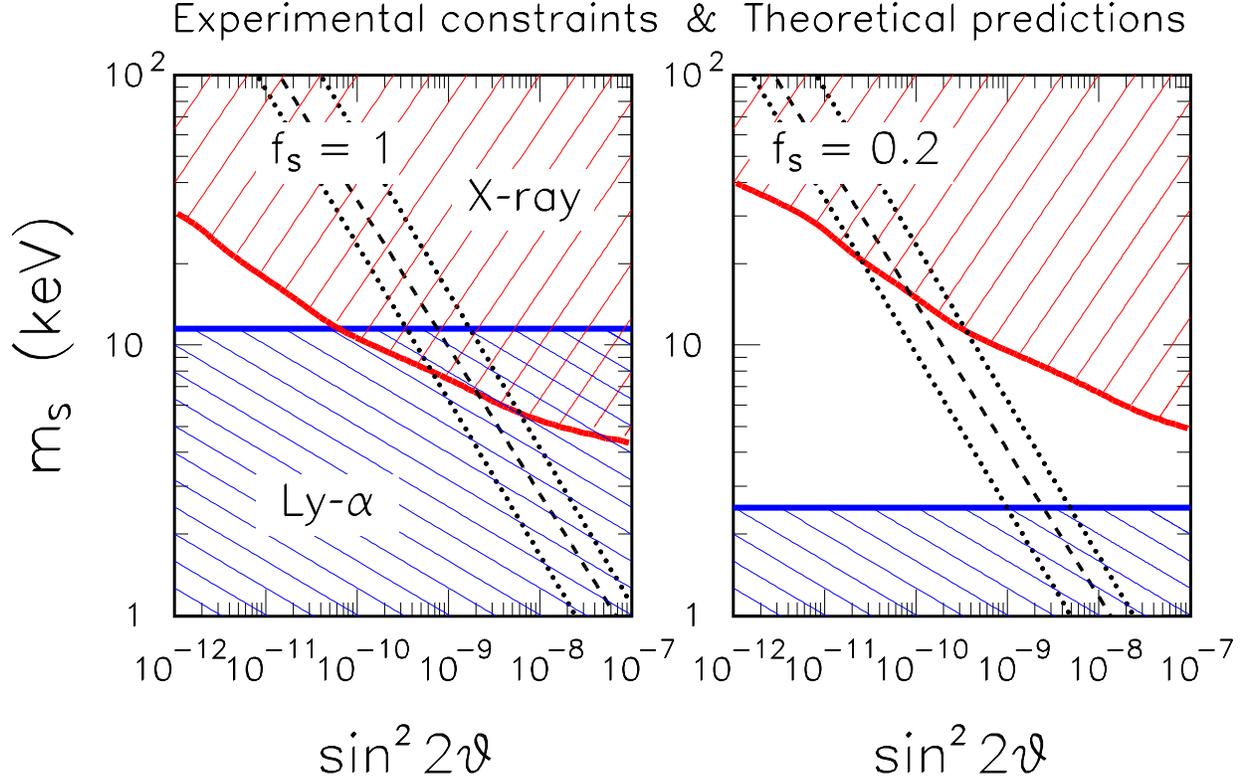}
\vspace{-4.0cm} \caption{\label{sections} Comparison of the
constraints on mass-mixing parameters imposed by experimental data
(X-ray and Ly-$\alpha$) with the theoretical predictions obtained
in~\cite{Asaka:2006rw} for the Dodelson-Widrow mechanism. The two
panels confront the ``usual'' case $f_s = 1$ (left panel) with a
representative case in which sterile neutrinos constitute only a
fraction $f_s = 0.2$ of dark matter (right panel).}
\end{figure}
%%%%%%%%%%%%%%%%%%%%%%%%%%%%%%%%%%%%%%%%%%%%%%%%%%%%%%%%%%%%%

%%%%%%%%%%%%%%%%%%%%%%%%%%%%%%%%%%%%%%%%%%%%%%%%%%%%%%%%%%%%%%%%%%%%%
\begin{figure}
\includegraphics[scale=0.90, bb= 100 100 510 720]{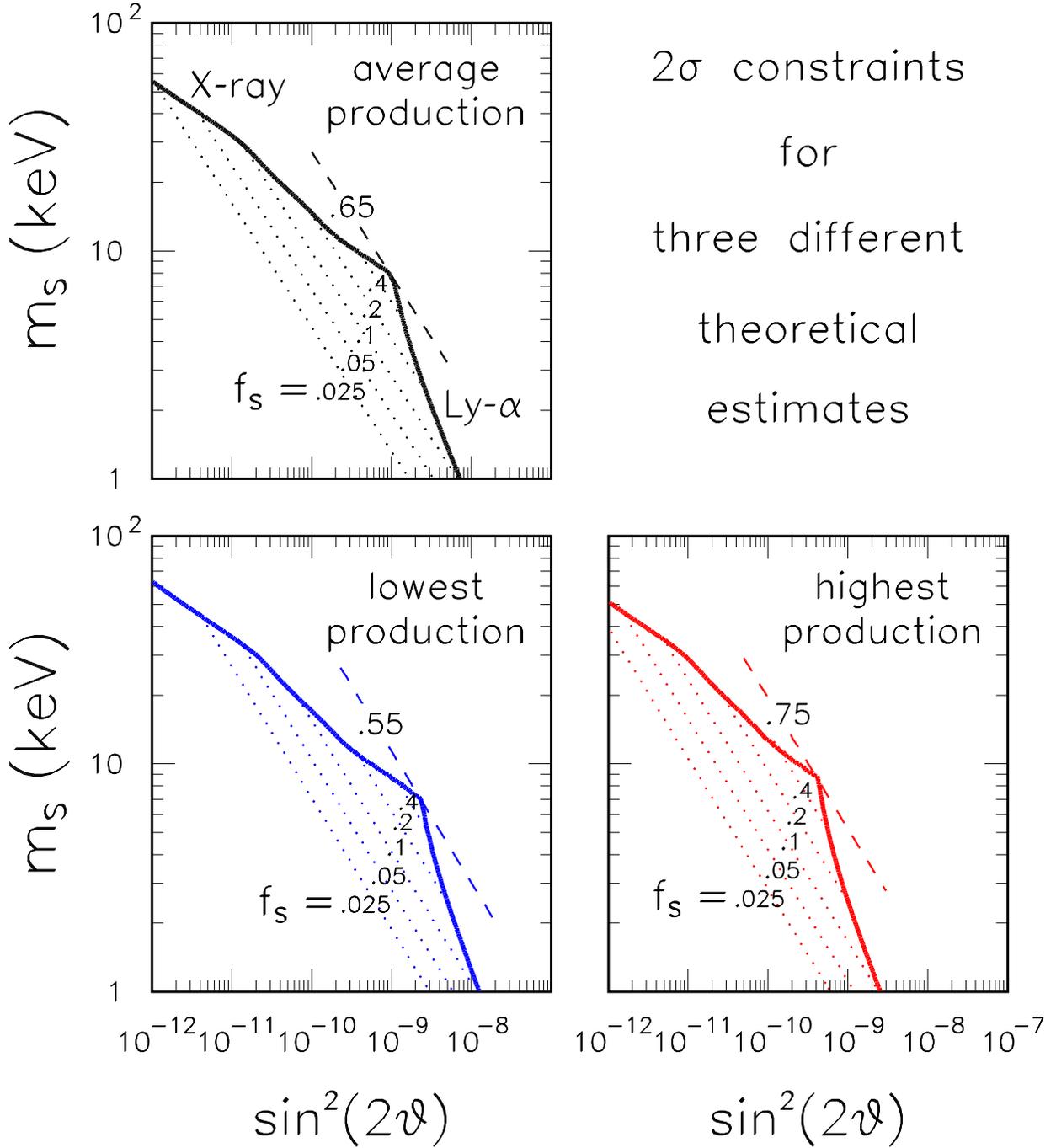}
\vspace{-0.0cm} \caption{\label{fig:qcd_impact} Joint constraints
from X-ray and Ly-$\alpha$ data on the mass-mixing parameters
obtained for three different theoretical estimates of the relic
abundance of sterile neutrinos produced via the Dodelson-Widrow
mechanism. In each panel the allowed region lies under the solid
curve, the dotted lines represent curves of constant sterile
neutrino fractional abundance, while the dashed segment indicates
its maximum value allowed in each case. In the first panel the
``average'' theoretical calculation obtained
in~\cite{Asaka:2006rw} is used. In the last two panels the
theoretical abundance is taken equal to the two extreme estimates
determined in~\cite{Asaka:2006rw}.}
\end{figure}
%%%%%%%%%%%%%%%%%%%%%%%%%%%%%%%%%%%%%%%%%%%%%%%%%%%%%%%%%%%%%%%%%%%%%

%%%%%%%%%%%%%%%%%%%%%%%%%%%%%%%%%%%%%%%%%%%%%%%%%%%%%%%%%%%%%%%%%%%%%
\begin{figure}
\includegraphics[scale=0.90, bb= 100 100 510 720]{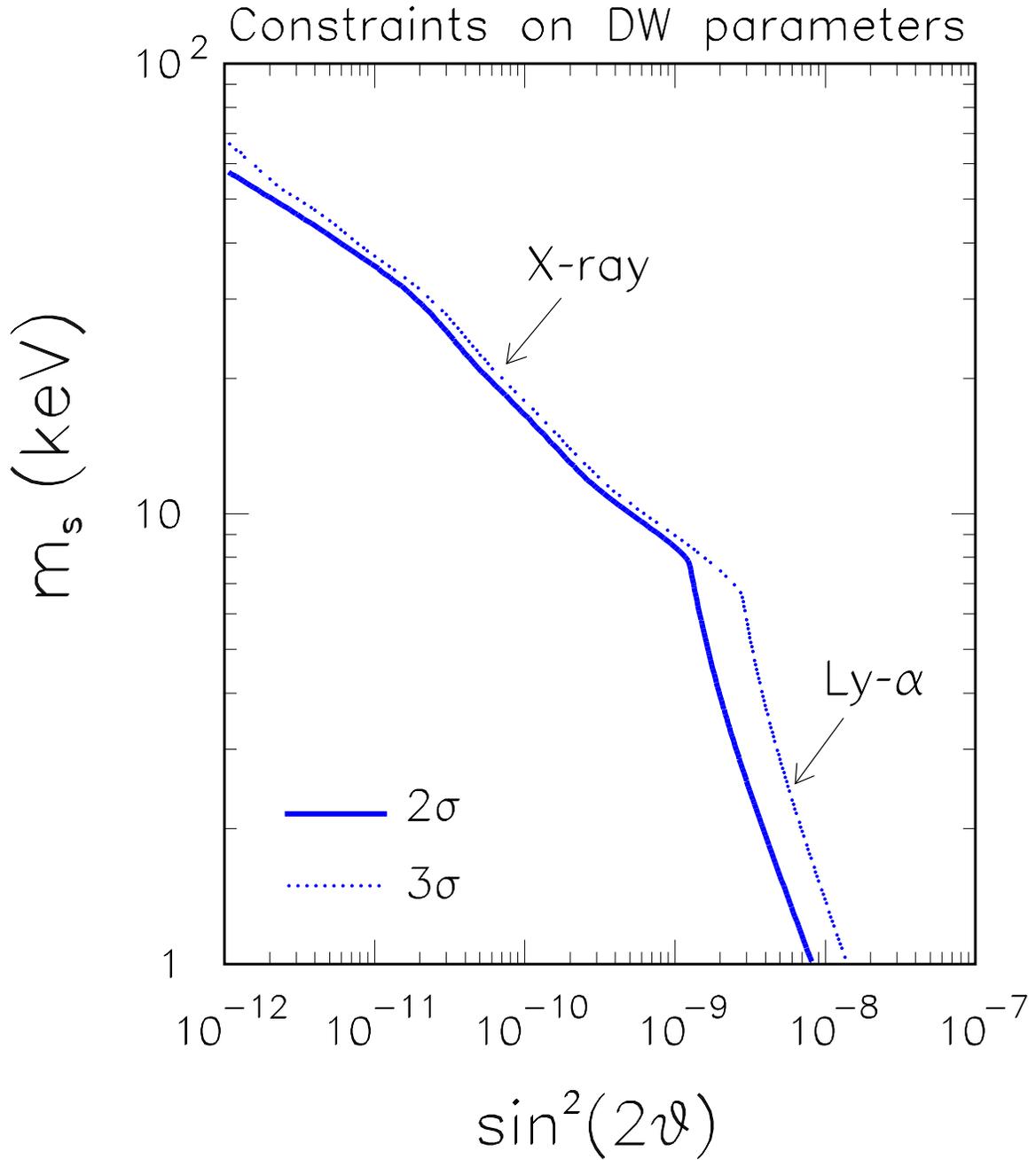}
\vspace{-0.0cm} \caption{\label{fig:mass_mixing} Constraints on
mass and mixing parameters governing the Dodelson-Widrow mechanism
obtained by the joint analysis of X-ray and Ly-$\alpha$ data with
the inclusion of the theoretical uncertainties. See the text for
details.}
\end{figure}
%%%%%%%%%%%%%%%%%%%%%%%%%%%%%%%%%%%%%%%%%%%%%%%%%%%%%%%%%%%%%%%%%%%%%

%%%%%%%%%%%%%%%%%%%%%%%%%%%%%%%%%%%%%%%%%%%%%%%%%%%%%%%%%%%%%%%%%%%%%
\begin{figure}
\includegraphics[scale=0.90, bb= 100 100 510 720]{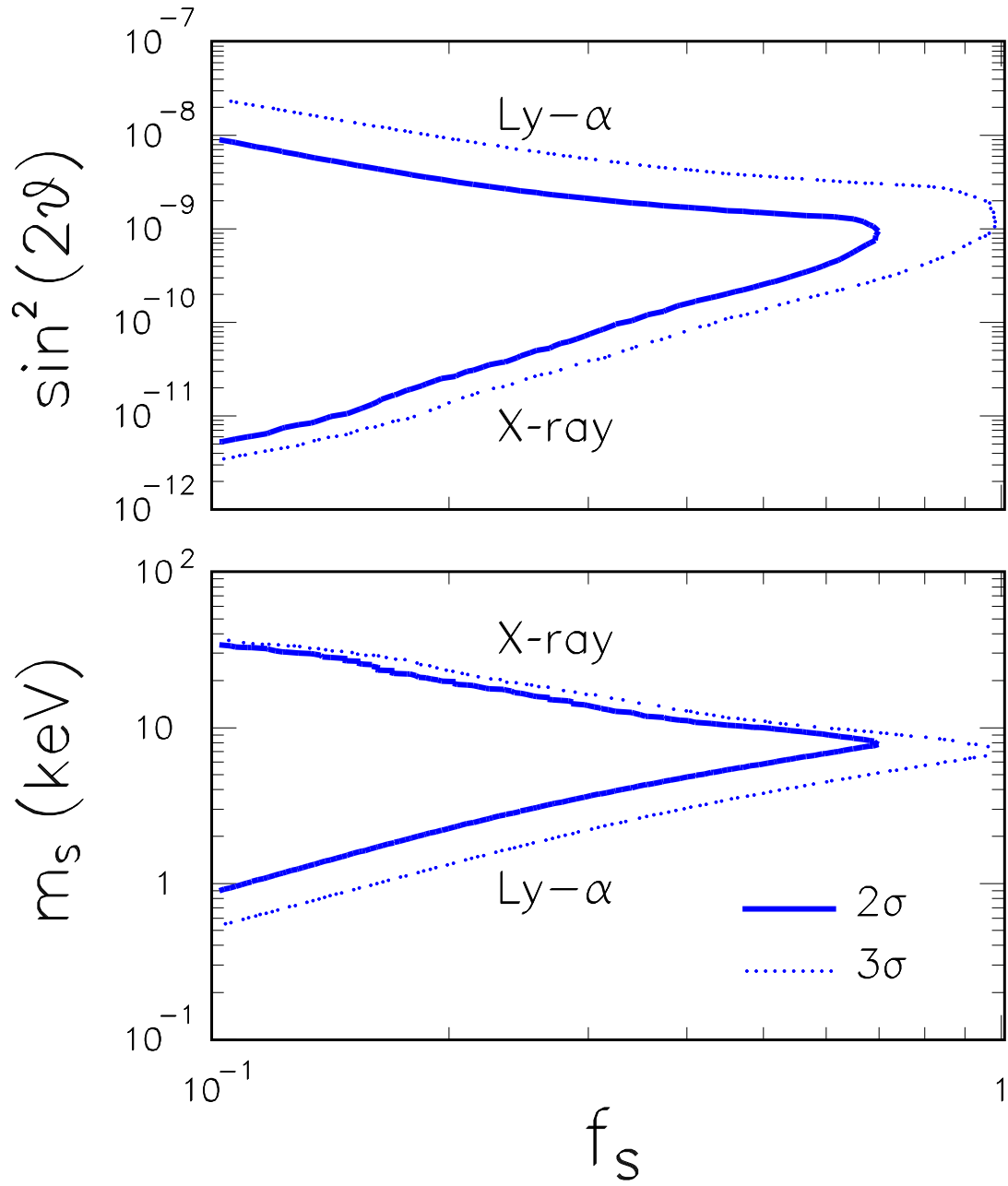}
\vspace{-0.0cm} \caption{\label{fig:ranges}  Ranges allowed for
the mixing angle (upper panel) and the neutrino mass (lower panel)
as a function of the relic abundance obtained under the assumption
of production via the Dodelson-Widrow mechanism.}
\end{figure}
%%%%%%%%%%%%%%%%%%%%%%%%%%%%%%%%%%%%%%%%%%%%%%%%%%%%%%%%%%%%%%%%%%%%%

\end{document}